\documentclass[sigconf, table, dvipsnames]{acmart}

\usepackage{booktabs} % For formal tables
\usepackage{xcolor}
\usepackage{pgfplots}
\pgfplotsset{width=8cm,compat=1.8}

\setcopyright{rightsretained}
\usepackage{float}
\usepackage{todonotes}

\usepackage[flushleft]{threeparttable}

\usepackage{multibib}
\newcites{R}{Selected papers}

\begin{document}

\copyrightyear{2019}
\acmYear{2019}
\setcopyright{acmlicensed}
\acmConference[ARES '19]{Proceedings of the 14th International Conference on Availability, Reliability and Security (ARES 2019)}{August 26--29, 2019}{Canterbury, United Kingdom}
\acmBooktitle{Proceedings of the 14th International Conference on Availability, Reliability and Security (ARES 2019) (ARES '19), August 26--29, 2019, Canterbury, United Kingdom}
\acmPrice{15.00}
\acmDOI{10.1145/3339252.3340329}
\acmISBN{978-1-4503-7164-3/19/08}

\title{Applying Security Testing Techniques to Automotive Engineering}
\titlenote{Produces the permission block, and
  copyright information}

\author{Irdin Pekaric}

\affiliation{%
  \institution{Department of Computer Science University of Innsbruck}
  \streetaddress{Technikerstraße 21A}
  \city{Innsbruck}
  \state{Austria}
  \postcode{A-6020}
}
\email{irdin.pekaric@uibk.ac.at}

\author{Clemens Sauerwein}

\affiliation{%
  \institution{Department of Computer Science University of Innsbruck}
  \streetaddress{Technikerstraße 21A}
  \city{Innsbruck}
  \state{Austria}
  \postcode{A-6020}
}
\email{clemens.sauerwein@uibk.ac.at}

\author{Michael Felderer}

\affiliation{%
  \institution{Department of Computer Science University of Innsbruck}
  \streetaddress{Technikerstraße 21A}
  \city{Innsbruck}
  \state{Austria}
  \postcode{A-6020}
}
\email{michael.felderer@uibk.ac.at}

\renewcommand{\shortauthors}{Pekaric et al.}

\begin{abstract}
Over the past few decades, the automotive industry was mostly focused on testing the safety aspects of a vehicle. However, this was not the case with security testing as it only began to be addressed recently. As a result, multiple approaches applying various security testing techniques on different software-based vehicle IT components emerged. With that said, the research and practice lack an overview about these techniques. In this paper, we conduct a systematic mapping study. This involved the investigation on the following five dimensions: (1) security testing techniques, (2) AUTOSAR layers, (3) functional interfaces of AUTOSAR, (4) vehicle lifecycle phases and (5) attacks. In total, 39 papers presenting approaches for security testing in automotive engineering were systematically selected and classified. The results identify multiple security testing techniques focusing on early phases of vehicle life cycle through the application and services layer of the AUTOSAR architecture. Finally, there is a need for security regression testing approaches, as well as combined security and safety testing approaches.      
\end{abstract}





\maketitle

\section{Introduction}
Originally, cars were designed as closed systems with a primary focus on functionality and safety, whereas modern cars are highly connected vehicles where security is becoming an increasingly important topic. This is made evident by an increasing number of attack vectors and corresponding attacks. For example, these attacks include remote exploitation of an unaltered passenger vehicle \cite{miller2015remote} or contactless attacks against sensors of self-driving vehicles \cite{yan2016can}. However, in contrast to safety testing, security testing techniques in the automotive engineering are in a very early stage~\cite{bayer2016automotive}. This is surprising since several researches outlined a strong need for security testing techniques \cite{sagstetter2013security} and there are increasing demands on automotive security made by public authorities \cite{markey2013wireless}.

Security testing techniques are well-established concepts in other fields, like software engineering~\cite{felderer2016sectesting}. These testing techniques include model-based testing, code-based testing, penetration testing and dynamic analysis, regressing testing and risk-based testing~\cite{felderer2016sectesting}. A few researchers adopted these testing techniques, like penetration testing \cite{durrwang2018enhancement} to the automotive domain. However, it is unclear to which extent these testing techniques are used in automotive engineering and how they relate to vehicle architectures and their lifecycle. Therefore, the research objective of this paper is to provide a comprehensive overview of applied security testing techniques in automotive engineering and to map them to the standard vehicle architecture AUTOSAR \cite{autosar} and the vehicle lifecycle of the standard ISO/SAE 21434 
\cite{iso}. It might become the standard for road vehicle cybersecurity engineering. In doing so we answer the following three research questions:

\begin{itemize}
	\item[RQ1] \textit{Which security testing techniques are applied in different stages of vehicle life-cycle?}
	\item[RQ2] \textit{Which attacks are addressed in different automotive architecture layers and security testing techniques?}
	\item[RQ3] \textit{How are different security testing techniques related to specific automotive architecture layers and functional interfaces?}
\end{itemize}
In order to address these research questions we conducted a systematic mapping study based on the methodology presented in \cite{petersen2015guidelines}. In doing so 207 relevant papers were analyzed which resulted, after inclusion and exclusion, in a final set of 39 papers. We analyzed this final set with regard to our three research question (RQ1 to RQ3). Thereby, we examined recent developments related to security testing approaches in the automotive domain and outlined the relationships between them. Finally, we mapped them to the AUTOSAR layers as a standardized vehicle architecture, attacks, functional interfaces of AUTOSAR and the vehicle lifecycle phases.

The related work in this field is very limited. We identified a systematic literature review on Internet-of-Vehicles communication security \cite{abu2018systematic}. However, the focus of this study is only on communication security, while we consider every layer of the AUTOSAR architecture.

The remainder of this paper is structured as follows: Section~\ref{sec:background} provides background information regarding security testing techniques in general and outlines the core concepts of the AUTOSAR architecture. Section~\ref{sec:methodology} discusses the research questions and describes the conducted systematic literature review. Section~\ref{sec:results} outlines the results of our mapping study. Section~\ref{sec:discussion} discusses the results and answers our research questions. Finally, Section~\ref{sec:conclusion} concludes the research at hand and provides outlook on future work.

\section{Background}
\label{sec:background}
In this section, we provide the necessary background information on \textit{security testing techniques} and the \textit{AUTOSAR} architecture. The provided information is used in this paper in order to classify publications based on the security testing techniques in the automotive domain.

\subsection{Security testing techniques}
\label{subsec:sectest}

\begin{figure}[h!]
	\centering
	\includegraphics[width=0.9\columnwidth]{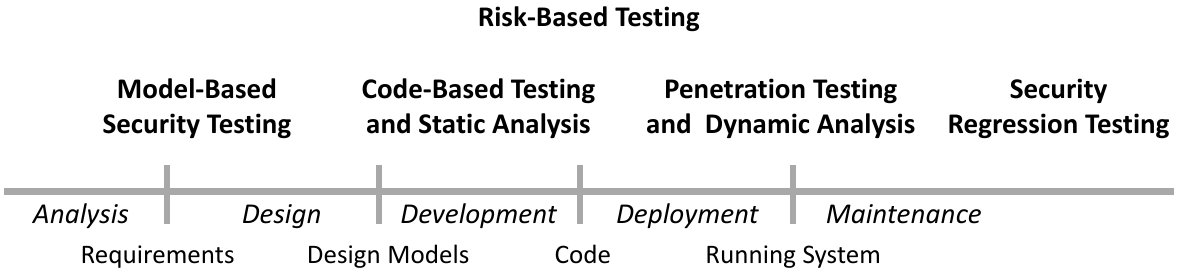}
	\caption{Security testing techniques in the secure system development lifecycle}
	\label{fig:sec_testing_activities}
	\centering
\end{figure}

Security testing covers (1) testing of security requirements that concerns confidentiality, integrity, availability, authentication, authorization, or non-repudiation and (2) testing to validate the ability of the software to withstand attack~\cite{cruzes2017security}. Security testing techniques therefore aim to ensure security functionality and to identify vulnerabilities. Based on \cite{felderer2016sectesting} security techniques can be classified according to their test basis within the secure system development lifecycle into the following five different types, which are also shown in Figure~\ref{fig:sec_testing_activities}:

\begin{itemize}
  \item \textit{Model-based testing} is grounded on requirements and design models created in the respective phases and applied to automatically generate security tests.
  \item \textit{Code-based testing and static analysis} is based on source and byte code that is white-box tested dynamically and statically during development.
  \item \textit{Penetration testing and dynamic analysis} is based on running systems, either in a test or production environment, to perform black-box testing or analysis of the system.
  \item \textit{Regression testing} ensures that changes applied to the system do not harm the security and is applied during maintenance.  
  \item \textit{Risk-based testing} guides security testing based on security risk analysis that relies on metrics from various artifacts from the secure development lifecycle like impact of requirements, code complexity or change rates. 
\end{itemize}

\subsection{AUTOSAR layers}
\label{subsec:autosarlayers}

\begin{figure}[h!]
	\centering
	\includegraphics[width=0.8\columnwidth]{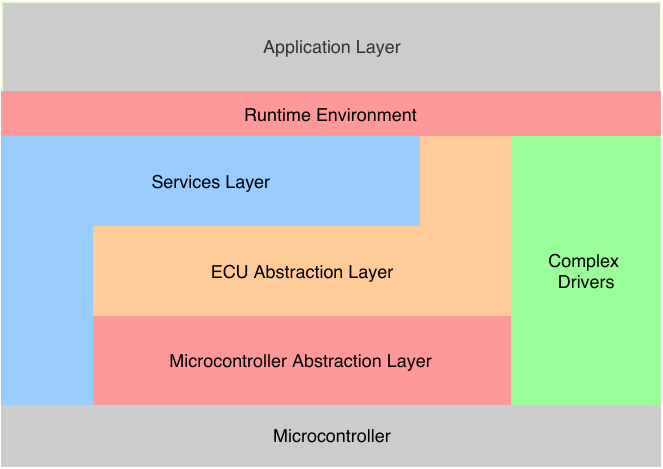}
	\caption{The AUTOSAR architecture \cite{autosar}}
	\label{fig:autosar_arch}
	\centering
\end{figure}

The AuTomotive Open System ARchitecture (AUTOSAR) is an open and standardized software architecture for electronic control units (ECUs) in the automotive domain (see Figure \ref{fig:autosar_arch}). The general AUTOSAR layered architecture consists of the following layers:
\begin{itemize}
  \item \textit{Application Layer}: Contains various types of applications that provide multiple functionalities and are executed depending on the use-case.
  \item \textit{Runtime Environment}: Provides communication services to the application software.
  \item \textit{Services Layer}: Offers operating system functionality, network communication, memory services, diagnostic services, ECU state management and program flow monitoring.
  \item \textit{ECU Abstraction Layer}: Provides an API in order to access peripherals and devices regardless of their location and connection to the operating system.
  \item \textit{Microcontroller Abstraction Layer}: Consists of internal drivers, allowing direct access to the operating system and internal peripherals.
  \item \textit{Complex Drivers}: Provide capability to integrate an additional functionality that is not specified within the AUTOSAR architecture.
  \item \textit{Microcontroller}: Runs all the aforementioned layers.
\end{itemize}

\section{Methodology}
\label{sec:methodology}

In order to obtain a comprehensive overview of security testing techniques used in the automotive domain, a systematic mapping study was conducted. In doing so, we applied the guidelines introduced by Petersen et al. \cite{petersen2015guidelines} and its refinement for security engineering by Felderer and Carver~\cite{felderer2018guidelines}.

In the following we demonstrate the \textit{search strategy}, \textit{search process}, \textit{selection criteria} and \textit{selection process}. Finally, we explain the \textit{data classification and analysis}, and address possible \textit{threats to validity}. 

\subsection{Search strategy}
The study was performed between December 2018 and January 2019. We based our literature review by following two different methodologies: a keyword search \cite{petersen2015guidelines} combined with forward and backward snowballing \cite{wohlin2014guidelines}. 

The keyword search was managed using Scopus \cite{Scopus}, which is an indexing database, allowing exploration of multiple online literature databases. As a result, we obtained an initial set of publications, which was used for conducting exhaustive forward (identification of new publications that cite the one being examined) and backward snowballing (examining the references of a publication being studied) iterations. This was done in order to obtain additional related publications and ensure the completeness of the final set. In this context, Kitchenham and Brereton \cite{kitchenham2013systematic} argue that applying multiple methodologies guarantees the sufficient literature coverage.

\subsection{Search process}

\begin{figure}[h!]
	\centering
	\includegraphics[width=0.35\textwidth]{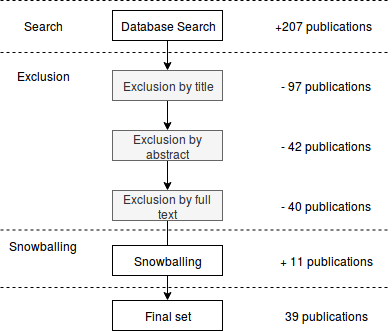}
	\caption{Search process}
	\label{fig:search_process}
	\centering
\end{figure}

The search process was conducted iteratively and consisted of: \textit{keyword identification}, \textit{database-search} and several \textit{snowballing iterations}. The whole process is illustrated in Figure \ref{fig:search_process}.

\textit{Keyword identification}. In order to identify the most appropriate search string, we conducted a preliminary search by experimenting using different sets of keywords. In doing so, we defined the following search string:  ("automotive"  OR  "car" )  AND  "security"  AND  "testing". We did not consider using "vehicle" as a keyword since it might include papers focusing on different types of vehicles that are not part of the automotive domain.  

\textit{Database search}. The database search was conducted using Scopus indexing database and the aforementioned search string. As a result, we obtained a set of 207 publications, which were retrieved based on the title, abstract and keywords. In order to create a good starting set, we applied inclusion/exclusion criteria (see Subsection \ref{subsec:selcriteria}) in the early stage of the search process.  

\textit{Snowballing}. The snowballing methodology was applied after the selection process was completed (see Section \ref{selection_process}). This was done to ensure that snowballing iterations were applied to a set of high quality publications. In doing so, we thought of five characteristics of a good start set defined by Wohlin et al. \cite{wohlin2014guidelines}. Therefore, we executed forward and backward snowballing on 28 papers until no new publications were found. 

\subsection{Selection criteria}
\label{subsec:selcriteria}

We selected papers based on the inclusion and exclusion criteria listed in Table \ref{tab:exclusion}. We considered only publications that were peer-reviewed in order to ensure high quality of the final set of papers. In addition, we addressed publications accessible only in full text. This is done in order to ensure the completeness of information allowing us to conduct a proper classification. Furthermore, we considered papers published between 2013 and 2019, allowing us to retrieve the most recent publications in the field. Finally, the last inclusion criteria denotes that we included only papers that present a security testing approach, which was the focus of this study.

On the other hand, we excluded gray and white literature (everything aside peer-reviewed conference, journal and workshop papers) ensuring that the resulting set is of high quality. Furthermore, non-English articles were excluded due to language constraints. In addition, we removed all the duplicate papers that were part of the search set. Since Scopus includes results from various libraries, some papers appeared more than once and they were excluded. Finally, we excluded all the papers focusing on general security topics and not on specific approaches as well as the papers that are outside of the automotive domain.

\begin{table}[]
	\centering
	\caption{Inclusion and exclusion criteria}
	\label{tab:exclusion}
	\begin{tabular}{|l|l|l|}
		\hline
		\multicolumn{2}{|l|}{\textit{\textbf{Inclusion Criteria}}}                                                                                             & \textit{\textbf{Exclusion Criteria}} \\ \hline
		\multicolumn{2}{|l|}{Peer reviewed articles}                                                                                                           & \textit{Gray and white literature}   \\ \hline
		\multicolumn{2}{|l|}{Accessible in Full text}                                                                                                          & \textit{Non-English articles}        \\ \hline
		\multicolumn{2}{|l|}{\begin{tabular}[c]{@{}l@{}}Published between 2013  \\ and 2019\end{tabular}}                                                      & Duplicates                           \\ \hline
		\multicolumn{2}{|l|}{\begin{tabular}[c]{@{}l@{}}Addressing Security\\ Testing Approaches \\ in Automotive Domain\end{tabular}} & \textit{\begin{tabular}[c]{@{}l@{}}General Security Topics \\ and Topics Outside \\ Automotive Domain\end{tabular}}                        \\ \hline
	\end{tabular}
\end{table}

\subsection{Selection process} \label{selection_process}
In addition to the stated selection criteria, we based the selection process on examining a publication's title, abstract and full text (see Figure \ref{fig:search_process}). This process was done iteratively. This means that the initial set of 207 publications was first evaluated based on the title. If the title suggested that the topic is not related to security testing technique, we removed the publication from the set. This resulted in exclusion of 97 papers. In the following step, we examined the resulting 110 papers based on the abstract. As a result, 42 additional publications were excluded. Finally, we conducted the full-text reading and excluded 40 more publications. As a result, the set of 28 publications remained (\citeR{alheeti2016intelligent}, \citeR{foggia2015car}, \citeR{park2016study}, \citeR{imparato2017comparative}, \citeR{abbott2016techniques}, \citeR{singh2015secure}, \citeR{wiersma2017safety}, \citeR{murvay2017attacks}, \citeR{mouttappa2013monitoring}, \citeR{wurzinger2016real}, \citeR{ji2018comparative}, \citeR{huang2018atg}, \citeR{cheah2016combining}, \citeR{macher2014combined}, \citeR{tomlinson2018detection}, \citeR{kurachi2018proposal}, \citeR{cheah2017formalising}, \citeR{corbett2017testing}, \citeR{groza2016evaluating}, \citeR{taylor2018probing}, \citeR{8416255}, \citeR{kong2018security}, \citeR{pike2017secure}, \citeR{humayed2017using}, \citeR{cheah2018building}, \citeR{kong2016security}, \citeR{bayer2015security}), which was used for executing the snowballing iterations. Finally, 11 new publications were identified with the snowballing methodology: (\citeR{7995934}, \citeR{inproceedings22}, \citeR{10.1007/978-3-319-13257-0_27}, \citeR{10.1371/journal.pone.0155781}, \citeR{6894181}, \citeR{CHEAH20178}, \citeR{Islam:2016:RAF:2899015.2899018}, \citeR{7796898}, \citeR{shoukry2013non}, \citeR{song2016intrusion}, \citeR{ruecker2013crash}), which were added to the final set of 39 papers.

\subsection{Classification and analysis}

The classification table (see Table \ref{tab:sel_pub}) was developed using the following five dimensions: (1) security testing techniques, (2) AUTOSAR layers, (3) functional interface of AUTOSAR, (4) vehicle lifecycle and (5) attack types. In the following, we discuss each of these categories with their concrete items in more detail.

\subsubsection{Security testing techniques} According to Section~\ref{subsec:sectest} we distinguish among the following security testing techniques: (1) \textit{model-based testing}, (2) \textit{code-based testing}, (3) \textit{penetration testing and dynamic analysis}, (4) \textit{regression testing}, and (5) \textit{risk-based testing}.

\subsubsection{AUTOSAR layers} According to Section~\ref{subsec:autosarlayers} AUTOSAR distinguishes among the following layers: (1) \textit{Application Layer}, (2) \textit{Runtime Environment}, (3) \textit{Services Layer}, (4) \textit{ECU Abstraction Layer}, (5) \textit{Microcontroller Abstraction Layer}, (6) \textit{Complex Drivers}, and (7)  \textit{Microcontroller}.

\subsubsection{AUTOSAR functional interface}
The AUTOSAR standardized multiple application interfaces  related to syntax and semantics. These are categorized into following six domains: (1) \textit{Body and comfort}, (2) \textit{Powertrain}, (3) \textit{Chasis}, (4) \textit{Safety}, (5) \textit{Multimedia/telematics}, and  (6) \textit{Man-machine-interface}. They are addressed in the technical overview of the AUTOSAR \cite{autosar}. In addition, we defined a new category for the AUTOSAR functional interfaces (\textit{not specified/general purpose}). This included all the security testing approaches that did not specify the targeted interface or the approach was applicable in general. 

\subsubsection{Vehicle lifecycle}
Similar to system-development lifecycle, a vehicle lifecycle represents the five phases of vehicle lifeline from design to decommissioning. These were derived from the upcoming ISO/SAE 21434 standard \cite{iso}. As a result, we consider the following five phases: (1) \textit{Design and engineering}, (2) \textit{Production}, (3) \textit{Operation by customer}, (4) \textit{Maintenance} and (5) \textit{Service and decommissioning}. 

\subsubsection{Attack types}
Initially, we extracted different types of attacks addressed in the papers. Considering that we encountered a large set of various attacks, we decided to cluster them in their respective groups. In doing so, we applied attack classification from the paper on threats of cyber attacks on critical infrastructures by Maglaras et al. \cite{maglaras2019threats}, allowing us to group all the specific attacks identified in the papers. Therefore, we differentiate between attacks focusing on: (1) \textit{Privacy}, (2) \textit{Integrity}, (3) \textit{Availability} and (4) \textit{Authentication}. We did not make a separate dimension for attacks on confidentiality because \cite{maglaras2019threats} considers them as attacks on privacy and authentication. For example, if a certain approach applied denial-of-service (DDoS) attacks, we would classify it as an attack targeting system's availability. 

Considering that a paper might address different aspects of each aforementioned category, it is possible to classify multiple dimensions. The papers were classified according to the procedure described in \cite{Carver:2016:EBM:2898375.2898380}, where the research team that conducted the classification consisted of a PhD student, a post-doctoral researcher and a faculty member. The classification was conducted by applying the following procedure. (1) Initially, the first author classified all the publications and recorded his results in a spreadsheet. In order to provide an accurate classification of papers, each dimension was discussed beforehand and defined by all three authors. (2) The classification was verified by the second and third author. They divided the list of sources in such a way, where the second author analyzed the first 20 papers and the third author analyzed the remaining 19 papers. Accordingly, their comments and results were recorded in the separate spreadsheet. (3) The classification discrepancies were marked for further analysis and discussion. (4) Finally, all three authors discussed any differences in the classification tables. If disagreements occurred, a majority vote was taken, which included all three authors. This resulted in each paper being classified by at least two authors. Finally, at the end of the procedure, we merged all the results into a single spreadsheet, which was used for deriving the answers to research questions.

\section{Results}
\label{sec:results}
This section presents demographics of the final set of papers and results of the analysis. Thereby, Section~\ref{sec:demographics} provides demographic information on the identified papers, while Section~\ref{sec:class} presents the answers to our research questions.

\begin{table*} 
\centering 
\caption{Selected publications} 
\begin{tabular}{l c @{\hspace{1.2\tabcolsep}} c @{\hspace{1.2\tabcolsep}} c @{\hspace{1.2\tabcolsep}} c @{\hspace{1.2\tabcolsep}} c @{\hspace{1.2\tabcolsep}}| @{\hspace{1.2\tabcolsep}} c @{\hspace{1.2\tabcolsep}} c @{\hspace{1.2\tabcolsep}} c @{\hspace{1.2\tabcolsep}} c @{\hspace{1.2\tabcolsep}} c @{\hspace{1.2\tabcolsep}} c @{\hspace{1.2\tabcolsep}} c @{\hspace{1.2\tabcolsep}}| @{\hspace{1.2\tabcolsep}} c @{\hspace{1.2\tabcolsep}} c @{\hspace{1.2\tabcolsep}} c @{\hspace{1.2\tabcolsep}} c @{\hspace{1.2\tabcolsep}} c @{\hspace{1.2\tabcolsep}} c @{\hspace{1.2\tabcolsep}} c @{\hspace{1.2\tabcolsep}} | @{\hspace{1.2\tabcolsep}} c @{\hspace{1.2\tabcolsep}} c @{\hspace{1.2\tabcolsep}} c @{\hspace{1.2\tabcolsep}} c @{\hspace{1.2\tabcolsep}} c @{\hspace{1.2\tabcolsep}}| @{\hspace{1.2\tabcolsep}} c @{\hspace{1.2\tabcolsep}} c @{\hspace{1.2\tabcolsep}} c @{\hspace{1.2\tabcolsep}} c @{\hspace{1.2\tabcolsep}}} 
\toprule 
& \multicolumn{5}{c @{\hspace{1.2\tabcolsep}}| @{\hspace{1.2\tabcolsep}}}{\textbf{STT}} & \multicolumn{7}{c @{\hspace{1.2\tabcolsep}}| @{\hspace{1.2\tabcolsep}}}{\textbf{AL}} &
\multicolumn{7}{c @{\hspace{1.2\tabcolsep}}|@{\hspace{1.2\tabcolsep}}}{\textbf{FIA}} &
\multicolumn{5}{c@{\hspace{1.2\tabcolsep}}|@{\hspace{1.2\tabcolsep}}}{\textbf{VLC}} &
\multicolumn{4}{@{\hspace{1.2\tabcolsep}}c}{\textbf{A}}\\ 
\textbf{P} & MB & CB & PT & RT & RB & AP & R & SE & E & MA & CD & MI & B & PO & CH & SA & MT & MM & N & DS & P & O & M & DE & PR & I & A & AU\\
\midrule
\citeR{cheah2017formalising} & \checkmark &  &  &  &  & \checkmark &  &  &  &  &  &  &  &  &  &  & \checkmark &  &  & \checkmark & \checkmark & \checkmark & &  & \checkmark & \checkmark & & \\
\citeR{foggia2015car} & \checkmark &  &  &  &  & \checkmark &  &  &  &  &  &  &  &  &  &  & \checkmark &  &  &  & \checkmark & \checkmark &  &  &  &  &  & \\
\citeR{macher2014combined} & \checkmark &  &  &  &  & \checkmark &  & \checkmark &  &  &  &  &  & \checkmark &  & \checkmark & \checkmark &  &  & \checkmark & \checkmark & \checkmark &  &  & \checkmark & \checkmark & \checkmark & \\
\citeR{cheah2016combining} & \checkmark &  &  &  &  & \checkmark &  & \checkmark &  &  &  &  &  &  &  &  & \checkmark &  &  & \checkmark & \checkmark & \checkmark &  &  & \checkmark &  & \checkmark & \\
\citeR{kurachi2018proposal} & \checkmark &  &  &  &  &  &  & \checkmark & \checkmark &  &  &  &  &  &  &  & \checkmark &  &  &  & \checkmark & \checkmark &  &  & \checkmark & \checkmark &  & \checkmark\\
\citeR{park2016study} & \checkmark &  &  &  &  &  &  & \checkmark &  &  &  &  &  &  &  &  & \checkmark &  &  & \checkmark &  &  &  &  & \checkmark &  &  & \\
\citeR{10.1007/978-3-319-13257-0_27} & \checkmark &  &  & &  &  &  & \checkmark &  &  &  &  &  &  &  &  & \checkmark & &  & \checkmark & \checkmark & \checkmark &  &  &  &  &  & \\
\citeR{inproceedings22} & \checkmark &  &  &  &  &  &  & \checkmark &  &  &  &  &  &  &  &  & \checkmark &  & & \checkmark &  &  &  &  &  & \checkmark &  & \\
\citeR{cheah2018building} & \checkmark &  & \checkmark &  &  & \checkmark &  & \checkmark & \checkmark &  &  &  &  &  &  &  & \checkmark &  &  &  & \checkmark &  &  &  & \checkmark &  & \checkmark & \checkmark\\
\citeR{CHEAH20178} & \checkmark &  & \checkmark & &  &  &  & \checkmark & \checkmark &  &  &  &  &  &  &  & \checkmark &  &  & \checkmark &  &  &  &  & \checkmark & \checkmark & \checkmark & \checkmark \\
\citeR{mouttappa2013monitoring} & \checkmark &  & \checkmark &  &  &  &  & \checkmark &  &  &  &  &  &  &  &  & \checkmark &  &  & \checkmark &  &  &  &  & \checkmark &  &  & \\
\citeR{kong2018security} & \checkmark &  &  &  & \checkmark & \checkmark &  & \checkmark & \checkmark &  &  &  &  & \checkmark &  & \checkmark & \checkmark & \checkmark &  & \checkmark & \checkmark & \checkmark &  &  & \checkmark & \checkmark &  & \\
\citeR{imparato2017comparative} &  & \checkmark &  &  &  & \checkmark &  & \checkmark &  &  &  &  &  &  &  &  &  &  & \checkmark & \checkmark &  &  &  &  &  &  &  & \\
\citeR{wiersma2017safety} &  & \checkmark &  &  &  &  &  &  &  &  &  & \checkmark &  &  &  & \checkmark &  &  &  & \checkmark &  &  &  &  &  &  & \checkmark & \\
\citeR{pike2017secure} &  & \checkmark & \checkmark &  &  &  &  & \checkmark &  &  &  &  &  &  &  &  & \checkmark &  &  & \checkmark &  &  &  &  & \checkmark &  &  & \checkmark\\
\citeR{bayer2015security} &  & \checkmark & \checkmark &  & \checkmark & \checkmark &  & \checkmark & \checkmark &  &  &  &  & \checkmark &  &  & \checkmark &  &  & \checkmark &  &  &  &  & \checkmark &  & \checkmark & \checkmark\\
\citeR{humayed2017using} &  &  & \checkmark &  &  &  &  & \checkmark & \checkmark &  &  &  &  & \checkmark &  &  &  &  &  & \checkmark &  &  &  &  &  &  & \checkmark & \\
\citeR{8416255} & & & \checkmark & &  &  &  & \checkmark & \checkmark & \checkmark &  &  & &  &  &  & \checkmark &  &  & \checkmark & \checkmark & \checkmark &  &  & \checkmark & \checkmark &  & \\
\citeR{taylor2018probing} &  &  & \checkmark &  &  &  &  & \checkmark & \checkmark &  &  &  &  &  &  &  & \checkmark &  &  & \checkmark & \checkmark & \checkmark &  &  &  &  & \checkmark & \\
\citeR{groza2016evaluating} &  &  & \checkmark &  &  &  &  &  &  & \checkmark &  & \checkmark &  &  &  &  &  &  & \checkmark & \checkmark &  &  &  &  & \checkmark & \checkmark &  & \\
\citeR{corbett2017testing} &  &  & \checkmark &  &  & \checkmark &  & \checkmark &  &  &  &  &  &  &  & & \checkmark &  &  & \checkmark & \checkmark & \checkmark &  &  & \checkmark & \checkmark &  & \checkmark\\
\citeR{tomlinson2018detection} &  &  & \checkmark &  &  &  &  & \checkmark & \checkmark &  &  &  &  &  &  &  & \checkmark &  &  &  & \checkmark & \checkmark &  &  &  & \checkmark &  & \\
\citeR{ji2018comparative} &  &  & \checkmark &  &  &  &  & \checkmark &  &  &  &  &  &  &  &  & \checkmark &  &  &  & \checkmark & \checkmark &  &  & \checkmark & \checkmark & \checkmark & \checkmark \\
\citeR{huang2018atg} &  &  & \checkmark &  &  &  &  & \checkmark &  &  &  &  &  &  &  &  & \checkmark &  &  & \checkmark &  &  &  &  & \checkmark & \checkmark & \checkmark & \\
\citeR{wurzinger2016real} &  &  & \checkmark &  &  & \checkmark &  & \checkmark & \checkmark &  &  &  &  &  &  &  & \checkmark &  &  & \checkmark & \checkmark & \checkmark &  &  & \checkmark &  & \checkmark & \\
\citeR{murvay2017attacks} &  &  & \checkmark &  &  & \checkmark &  & \checkmark &  &  &  &  &  &  &  &  & \checkmark &  &  & \checkmark &  &  &  &  &  &  & \checkmark & \\
\citeR{singh2015secure} &  &  & \checkmark &  &  &  &  & \checkmark & &  &  &  &  &  &  &  & \checkmark &  &  &  &  & \checkmark &  &  &  &  & \checkmark & \\
\citeR{abbott2016techniques} &  &  & \checkmark &  &  &  &  & \checkmark & \checkmark &  &  &  &  &  &  &  & \checkmark &  &  & \checkmark &  &  &  &  &  &  & \checkmark & \\
\citeR{alheeti2016intelligent} &  & & \checkmark &  &  &  &  &  &  &  & \checkmark &  &  &  &  &  & \checkmark &  &  & \checkmark &  &  &  &  &  &  &  & \\
\citeR{ruecker2013crash} &  &  & \checkmark &  &  &  &  & \checkmark &  &  &  &  & \checkmark &  &  &  &  &  &  &  &  & \checkmark &  &  &  &  &  & \\
\citeR{song2016intrusion} &  &  & \checkmark &  &  &  &  & \checkmark &  &  &  &  &  &  &  &  & \checkmark &  &  & \checkmark &  &  &  &  &  &  & \checkmark & \\
\citeR{shoukry2013non} &  &  & \checkmark &  &  & \checkmark &  & \checkmark &  &  & \checkmark &  &  &  &  &  & \checkmark &  &  & \checkmark &  &  &  &  & \checkmark &  &  & \\
\citeR{7796898} &  &  & \checkmark &  &  &  &  & \checkmark &  &  &  &  &  &  &  &  & \checkmark &  &  & \checkmark &  &  &  &  &  &  & \checkmark & \\
\citeR{6894181} &  &  & \checkmark &  &  &  &  & \checkmark &  &  &  &  &  &  &  &  & \checkmark & &  & \checkmark &  &  &  &  & \checkmark &  &  & \\
\citeR{10.1371/journal.pone.0155781} &  &  & \checkmark &  &  &  & & \checkmark &  &  &  &  &  &  &  &  & \checkmark &  &  & \checkmark & \checkmark & \checkmark &  &  &  & \checkmark &  & \\
\citeR{article} &  &  & \checkmark &  &  &  & & \checkmark &  &  &  &  & &  &  &  & \checkmark &  &  & \checkmark & \checkmark &  &  &  &  & &  & \\
\citeR{7995934} &  &  & \checkmark &  &  &  &  & \checkmark &  &  &  &  &  & &  &  & \checkmark &  &  & \checkmark & \checkmark & \checkmark &  &  &  & \checkmark &  & \\
\citeR{Islam:2016:RAF:2899015.2899018} &  &  &  & & \checkmark & \checkmark &  & \checkmark &  &  &  &  &  &  &  &  & \checkmark &  &  &  & \checkmark &  &  &  & \checkmark & \checkmark & \checkmark & \checkmark \\
\citeR{kong2016security} &  &  &  &  & \checkmark &  &  & \checkmark & \checkmark &  &  &  &  &  &  & \checkmark & \checkmark &  &  &  & \checkmark & \checkmark &  &  &  &  &  & \\  
%\midrule 
%\midrule
%SUM & 12 & 4 & 26 & 0 & 4 & 1 & 1 & 1 & 1 & 1 & 1 & 1 & 1 & 1 & 1 & 1 & 1 & 1 & 1 & 1 & 1 & 1 & 1 & 1 & 1 & 1 & 1 & 1\\ 
\bottomrule 
\end{tabular}
\begin{tablenotes}
      \small
      \item STT = Security Testing Techniques (MB = Model-Based Testing, CB = Code-Based Testing, PT = Penetration Testing and Dynamic Analysis, RT = Regression Testing, RB = Risk Based Testing), AL = AUTOSAR Layers (AP = Application, R = Runtime Environment, SE = Services, E = ECU Abstraction, MA = Microcontroller Abstraction, CD = Complex Drivers, MI = Microcontroller), FIA = Functional Interfaces of AUTOSAR (B = Body/Comfort, PO = Powertrain, CH = Chasis, SA = Safety, MT = Multimedia/Telematics, MM = Man-Machine-Interface, N = Not Specified/General), VLC = Vehicle Life Cycle (DS = Design and Engineering, P = Production, O = Operation by Customer, M = Maintenance, DE = Decommissioning), A = Attacks (PR = Privacy, I = Integrity, A = Availability, AU = Authorization)
\end{tablenotes}
\label{tab:sel_pub} 
\end{table*}

\subsection{Demographics}
\label{sec:demographics}
In this section we present demographics related to the final set of selected papers. This includes information such as the frequency of publications, type of source and venue type. 
%\begin{figure}[h!]
%	\centering
%	\includegraphics[width=0.45\textwidth]{img/frequency_publications.pdf}
%	\caption{Frequency of publications}
%	\label{fig:f_publications}
%	\centering
%\end{figure}
 
\begin{figure}[h!]
\caption{Frequency of publications per venue type}
\label{fig:freq_venue}
\begin{tikzpicture}
\begin{axis}[
    ybar stacked,
	bar width=15pt,
	nodes near coords,
    enlargelimits=0.15,
    legend style={at={(0.5,-0.20)},
      anchor=north,legend columns=-1},
    ylabel={\#participants},
    symbolic x coords={2013,2014,2015,2016,2017,2018},
    xtick=data,
    x tick label style={rotate=45,anchor=east},
    ]
\addplot+[ybar] plot coordinates {(2013,0) (2014,2) 
  (2015,2) (2016,9) (2017,7) (2018,4) };
\addplot+[ybar] plot coordinates {(2013,1) (2014,0) 
  (2015,0) (2016,2) (2017,2) (2018,5) };
\addplot+[ybar] plot coordinates {(2013,2) (2014,0)
  (2015,0) (2016,1) (2017,2) (2018,0) };
\legend{\strut conference, \strut journal, \strut workshop}
\end{axis}
\end{tikzpicture}
\end{figure}
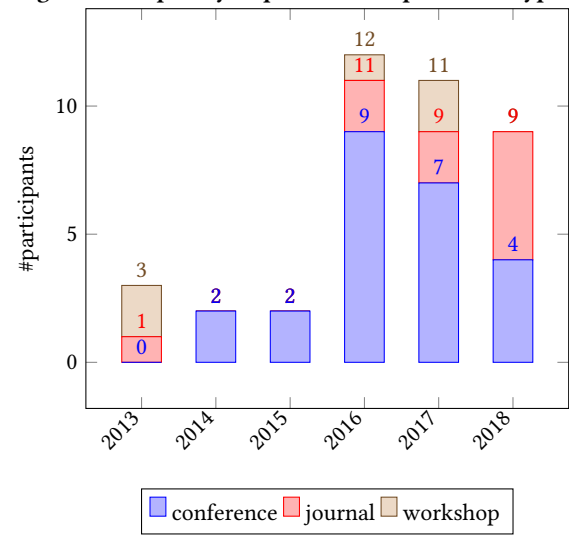
Figure \ref{fig:freq_venue} depicts the number of publications on security testing approaches in the automotive domain per venue type between the years 2013-2018. We observe a significant increase in the last three years, where the number of publications increased up to six times compared to the previous years. For example, in 2013 there were two papers, while in 2016 we identified 12 publications. This indicates that in the past couple of years, researchers are considering security testing in this domain to a higher extent. The slight decrease in 2018 and no publications in 2019 can be related to a period when this study was conducted (see Section 3.5). Thus, it is possible that some publications that were published at the end of 2018 and beginning of 2019 were not included in the final set. 

In addition, we consider publications based on the venue type. These include only peer-reviewed venues. Thus, we differentiate between \textit{workshop}, \textit{journal} and \textit{conference}. The most of the publications are published conferences (24). On the other hand, the amount of publications in journals (10) and workshops (5) is comparably smaller. This indicates that the papers in this field are most often submitted to conferences. However, in the last three years, we notice an increased number of journal publications compared to the previous years.  

%\noindent
%\begin{table}[ht]\caption{Publications per type of source}
%\setlength{\tabcolsep}{20pt}
%\begin{tabular}{@{} *5l @{}}    \toprule
%\emph{Type of source} & \emph{Number of publications} &  \\\midrule
%Industry    & 8    \\ 
% Combined  & 13  \\ 
% Academia  & 18  \\\bottomrule
% \hline
%\end{tabular}
% \label{table:tsource}
%\end{table}
Finally, we considered publications based on the type of a source. In doing so, we differentiated between authors from \textit{academia}, \textit{industry} and \textit{combined}. We define \textit{combined} as authors of a paper that come from both industry and academia. Based on this, it is possible to know the degree of cooperation between the fields. We identified eight industry, 13 combined and 18 academia papers. This is surprising because we expected more industry papers as this is the automotive domain. 
%\noindent
%\begin{table}[ht]\caption{Publications per venue type}
%\setlength{\tabcolsep}{20pt}
%\begin{tabular}{@{} *5l @{}}    \toprule
%\emph{Venue type} & \emph{Number of publications} &  \\\midrule
%Workshop    & 5    \\ 
% Journal  & 10  \\ 
% Conference  & 24  \\\bottomrule
% \hline
%\end{tabular}
% \label{table:tvenue}
%\end{table}

\definecolor{G1}{rgb}{216,216,216}
\definecolor{G2}{rgb}{176,176,176}
\definecolor{G3}{rgb}{128,128,128}

\begin{table*}[ht]
\centering
\caption{Security testing technique (STT) per vehicle lifecycle phase (VLP)}
\label{fig:t_lifecycle}
\begin{tabular}[t]{lccccc}
\toprule
STT/VLP & Design and Engineering & Production & Operation by Customer & Maintenence and Service & Decommissioning\\
\midrule
Model-based&\cellcolor[RGB]{176,176,176}9&\cellcolor[RGB]{176,176,176}8&\cellcolor[RGB]{176,176,176}7&0&0\\
Code-based&\cellcolor[RGB]{216,216,216}4&0&0&0&0\\
PTDA &\cellcolor[RGB]{128,128,128}21&\cellcolor[RGB]{176,176,176}10&\cellcolor[RGB]{176,176,176}10&0&0\\
Regression Testing &\cellcolor{white}0&0&0&0&0\\
Risk-based Testing  &\cellcolor[RGB]{216,216,216}2&\cellcolor[RGB]{216,216,216}3&\cellcolor[RGB]{216,216,216}2&0&0\\
\bottomrule
\end{tabular}
\end{table*}%

\begin{table}[ht]
\centering
\caption{Security testing technique per attack type}
\label{fig:t_attacks}
\begin{tabular}[t]{l@{\hspace{1.2\tabcolsep}}c@{\hspace{1\tabcolsep}}c@{\hspace{1\tabcolsep}}c@{\hspace{1\tabcolsep}}c@{\hspace{1\tabcolsep}}}
\toprule
STT/Attacks & Privacy & Integrity & Availability & Authenticity \\
\midrule
Model-based&\cellcolor[RGB]{176,176,176}9&\cellcolor[RGB]{176,176,176}6&\cellcolor[RGB]{216,216,216}4&\cellcolor[RGB]{216,216,216}3\\
Code-based&\cellcolor[RGB]{216,216,216}2&0&\cellcolor[RGB]{216,216,216}2&\cellcolor[RGB]{216,216,216}2\\
PTDA &\cellcolor[RGB]{128,128,128}13&\cellcolor[RGB]{176,176,176}9&\cellcolor[RGB]{128,128,128}13&\cellcolor[RGB]{176,176,176}6\\
Regression Testing &\cellcolor{white}0&0&0&0\\
Risk-based Testing  &\cellcolor[RGB]{216,216,216}3&\cellcolor[RGB]{216,216,216}2&\cellcolor[RGB]{216,216,216}2&\cellcolor[RGB]{216,216,216}2\\
\bottomrule
\end{tabular}
\end{table}%

\subsection{Classification of security testing techniques in the automotive domain}
\label{sec:class}
We conducted a classification of 39 selected papers based on the security techniques that were applied in their respective approaches (i.e. model-based, code-based, penetration testing and dynamic analysis, regression and risk-based). In this section we provide answers on the research questions (RQ 1-3) defined in Section 1.

\subsubsection{Security testing techniques to vehicle lifecycle}
Table \ref{fig:t_lifecycle} depicts a heat-map of security testing techniques per vehicle lifecycle phase. The focus of investigated approaches is on the first three phases of vehicle lifecycle (design and engineering \citeR{song2016intrusion} \citeR{7796898}, production \citeR{CHEAH20178} and operation by customer \citeR{ruecker2013crash}). This means that security issues are not considered at all in the last two phases (maintenance and service and decommissioning). 

With regards to security testing techniques, the results state that the most used technique is the penetration testing and dynamic analysis, which is applied in 26 selected publications \citeR{corbett2017testing} \citeR{abbott2016techniques}. This is followed by the model-based (12 papers) \citeR{cheah2018building} \citeR{cheah2016combining}, code-based (4) \citeR{imparato2017comparative} \citeR{wiersma2017safety} and risk-based (4) \citeR{kong2016security} \citeR{Islam:2016:RAF:2899015.2899018} testing. On the other hand, we did not encounter any approaches focusing on regression testing. This indicates that security testing approaches in the automotive domain focus on evaluating vehicles before their delivery to customers.

In addition, we analyzed the relation between security testing techniques and vehicle lifecycle phases. The results indicate a strong relation between \textit{penetration testing and dynamic analysis testing} techniques and \textit{design and engineering} lifecycle phase (20). In addition, we observed a relation between the \textit{penetration testing and dynamic analysis} and \textit{production} phase (10), \textit{penetration testing and dynamic analysis} and \textit{operation by customer} phase (10), \textit{model-based testing} and \textit{design and engineering} phase (9), \textit{model-based testing} and \textit{production} phase (8) and \textit{model-based testing} and \textit{operation by customer} phase (7). \textit{Risk-based testing} approaches are rather evenly spread through the first three phases of lifecycle (2,3,2). Finally, there are no approaches addressing \textit{code-based testing} in \textit{production} and \textit{operation by customer} phases, as well as no regression testing approaches in any of the phases of vehicle lifecycle. 

%\begin{table*}[ht]
%\centering
%\caption{Thicker horizontal lines above and below the table.}
%\begin{tabular}[t]{lccccc}
%\toprule
%STT/VLP & Design and Engineering & Production & Operation by Customer & Maintenence and Service & Decommissioning\\
%\midrule
%Model-based&\cellcolor{YellowOrange}9&\cellcolor{YellowOrange}8&\cellcolor{YellowOrange}7&\cellcolor{ForestGreen}0&\cellcolor{ForestGreen}0\\
%Code-based&\cellcolor{Goldenrod}4&\cellcolor{ForestGreen}0&\cellcolor{ForestGreen}0&\cellcolor{ForestGreen}0&\cellcolor{ForestGreen}0\\
%PTDA &\cellcolor{RedOrange}21&\cellcolor{YellowOrange}10&\cellcolor{YellowOrange}10&\cellcolor{ForestGreen}0&\cellcolor{ForestGreen}0\\
%Regression Testing &\cellcolor{ForestGreen}0&\cellcolor{ForestGreen}0&\cellcolor{ForestGreen}0&\cellcolor{ForestGreen}0&\cellcolor{ForestGreen}0\\
%Risk-based Testing  &\cellcolor{Goldenrod}2&\cellcolor{Goldenrod}3&\cellcolor{Goldenrod}2&\cellcolor{ForestGreen}0&\cellcolor{ForestGreen}0\\
%\bottomrule
%\end{tabular}
%\end{table*}%

%\begin{figure*}[h!]
%	\centering
%	\includegraphics[width=0.7\textwidth]{img/t_lifecycle.png}
%	\caption{Security testing technique per vehicle lifecycle phase}
%	\label{fig:t_lifecycle}
%	\centering
%\end{figure*}

\begin{table*}[ht]
\centering
\caption{Attack type per AUTOSAR layer (AUTOSAR L.)}
\label{fig:a_AUTOSAR}
\begin{tabular}[t]{lccccccc}
\toprule
Attack/AUTOSAR L & Application & \shortstack{Runtime \\ Environment} & Services & \shortstack{ECU Abstraction \\ Layer} & \shortstack{Microcontroller \\ Abstraction Layer} & \shortstack{Complex \\ Drivers} & Microcontroller\\
\midrule
Privacy&\cellcolor[RGB]{176,176,176}10&0&\cellcolor[RGB]{128,128,128}18&\cellcolor[RGB]{176,176,176}7&\cellcolor[RGB]{216,216,216}2&\cellcolor[RGB]{216,216,216}1&\cellcolor[RGB]{216,216,216}1\\
Integrity&\cellcolor[RGB]{216,216,216}5&0&\cellcolor[RGB]{128,128,128}13&\cellcolor[RGB]{216,216,216}5&\cellcolor[RGB]{216,216,216}2&0&\cellcolor[RGB]{216,216,216}1\\
Availability &\cellcolor[RGB]{176,176,176}7&0&\cellcolor[RGB]{128,128,128}15&\cellcolor[RGB]{176,176,176}7&0&0&\cellcolor[RGB]{216,216,216}1\\
Authenticity &\cellcolor[RGB]{216,216,216}4&0&\cellcolor[RGB]{176,176,176}8&\cellcolor[RGB]{216,216,216}4&0&0&0\\
\bottomrule
\end{tabular}
\end{table*}%

\begin{table*}[ht]
\centering
\caption{Security testing technique per AUTOSAR layer}
\label{fig:t_AUTOSAR}
\begin{tabular}[t]{lccccccc}
\toprule
STT/AUTOSAR L & Application & \shortstack{Runtime \\ Environment} & Services & \shortstack{ECU Abstraction \\ Layer} & \shortstack{Microcontroller \\ Abstraction Layer} & \shortstack{Complex \\ Drivers} & Microcontroller\\
\midrule
Model-based&\cellcolor[RGB]{176,176,176}6&0&\cellcolor[RGB]{176,176,176}10&\cellcolor[RGB]{216,216,216}4&0&0&0\\
Code-based&\cellcolor[RGB]{216,216,216}2&0&\cellcolor[RGB]{216,216,216}3&\cellcolor[RGB]{216,216,216}1&0&0&\cellcolor[RGB]{216,216,216}1\\
PTDA &\cellcolor[RGB]{176,176,176}6&0&\cellcolor[RGB]{128,128,128}24&\cellcolor[RGB]{176,176,176}9&\cellcolor[RGB]{216,216,216}2&\cellcolor[RGB]{216,216,216}2&\cellcolor[RGB]{216,216,216}1\\
Regression Testing &\cellcolor{white}0&0&0&0&0&0&0\\
Risk-based Testing  &\cellcolor[RGB]{216,216,216}3&0&\cellcolor[RGB]{216,216,216}4&\cellcolor[RGB]{216,216,216}2&0&0&0\\
\bottomrule
\end{tabular}
\end{table*}%

\begin{table*}[ht]
\centering
\caption{Security testing technique per functional interface of AUTOSAR (FI AUTOSAR)}
\label{fig:t_interface}
\begin{tabular}[t]{lccccccc}
\toprule
STT/FI AUTOSAR & \shortstack{Body/\\ Comfort } & \shortstack{Powertrain} & Chassis & \shortstack{Safety} & \shortstack{Multimedia/ \\ Telematics} & \shortstack{Man-Machine-Interface} & \shortstack{Not Specified/ \\ General Purpose} \\
\midrule
Model-based&0&\cellcolor[RGB]{216,216,216}2&0&\cellcolor[RGB]{216,216,216}2&\cellcolor[RGB]{176,176,176}12&\cellcolor[RGB]{216,216,216}1&0\\
Code-based&0&\cellcolor[RGB]{216,216,216}1&0&\cellcolor[RGB]{216,216,216}1&\cellcolor[RGB]{216,216,216}2&0&\cellcolor[RGB]{216,216,216}1\\
PTDA &\cellcolor[RGB]{216,216,216}1&\cellcolor[RGB]{216,216,216}2&0&\cellcolor[RGB]{216,216,216}1&\cellcolor[RGB]{128,128,128}22&0&\cellcolor[RGB]{216,216,216}1\\
Regression Testing &\cellcolor{white}0&0&0&0&0&0&0\\
Risk-based Testing  &0&\cellcolor[RGB]{216,216,216}2&0&\cellcolor[RGB]{216,216,216}2&\cellcolor[RGB]{216,216,216}4&\cellcolor[RGB]{216,216,216}1&0\\
\bottomrule
\end{tabular}
\end{table*}%

\subsubsection{Attacks to AUTOSAR layers}
During the analysis of approaches presented in the selected publications, we encountered that multiple security testing approaches apply to multiple attacks. Thus, we decided to investigate the relationship between attacks, security testing techniques and AUTOSAR layers (see Table \ref{fig:t_attacks} and Table \ref{fig:a_AUTOSAR}). In doing so, we identified that attacks focusing on privacy and availability are conducted more often compared to attacks focusing on integrity and authentication. On the other hand, the most addressed AUTOSAR layers are application, services and ECU abstraction layer. With relation to the specific attacks, we identified man-in-the-middle \citeR{cheah2016combining}, spoofing \citeR{macher2014combined}, sniffing \citeR{park2016study}, eavesdropping \citeR{ji2018comparative}, message modification \citeR{groza2016evaluating}, denial of service \citeR{cheah2018building}, insider \citeR{bayer2015security}, fault injection \citeR{bayer2015security}, fuzzing \citeR{CHEAH20178} and privilege escalation attacks \citeR{Islam:2016:RAF:2899015.2899018}.

Furthermore, we identified 13 penetration testing and dynamic analysis techniques applying attacks aiming to disrupt privacy and availability, nine on integrity and six on authentication. In addition, we recorded that model-based techniques in majority of cases apply attacks on system's privacy (9) and integrity (6). In code-based testing, we identified two publications using attacks on privacy, two on availability and two on authentication. However, we did not encounter any approaches using attacks on integrity together with the code-based testing. Finally, in the approaches applying risk-based testing, we discovered that attacks from each group were identified.

In regards to relation between attacks and AUTOSAR layers, we identified that attacks from each group were used on application, services and ECU abstraction layer. As a result, we discovered that approaches applying attacks on privacy (18) and availability (15) were addressing the service layer to a high extent. On the other hand, there were no attacks focusing on availability and authentication in the micro controller abstraction layer, as well as no attacks addressing integrity, availability and authentication in the complex drivers layer. 

\subsubsection{Security testing techniques to AUTOSAR layers and functional interfaces of AUTOSAR}
In order to investigate the application of different security testing techniques to AUTOSAR layers and functional interfaces of AUTOSAR, we map their relationships using the heatmap. As depicted in Table \ref{fig:t_AUTOSAR}, most of the security testing approaches focus on service, application and ECU abstraction layer of AUTOSAR architecture. As a result, we highlight 24 approaches performing penetration testing and dynamic analysis on the services layer. In addition, 10 model-based techniques focus on the services layer of AUTOSAR architecture, nine penetration testing and dynamic analysis techniques on the ECU abstraction layer, six model-based and six penetration testing and dynamic analysis techniques on the application layer. Furthermore, one code-based and one penetration testing and dynamic analysis technique apply to the microcontroller layer. Finally, penetration testing and dynamic analysis is the only technique used on the microcontroller abstraction and complex drivers layer.

With regards to the relationship between security testing techniques and functional interfaces of AUTOSAR, we observed a high focus on multimedia/telematics interface. As previously mentioned, this is due to many approaches targeting CAN bus for security testing. Results show 22 penetration testing and dynamic analysis, and 12 model-based approaches targeting multimedia/telematics interface. In addition, we observed a modest use of security testing techniques on the powertrain and safety functional interfaces. The remaining set of functional interfaces is merely addressed or not at all. This includes the body/comfort, chasis, man-machine-interface and 'not specified/general purpose' interfaces.
%\begin{figure*}[h!]
%	\centering
%	\includegraphics[width=0.7\textwidth]{img/t_AUTOSAR.png}
%	\caption{Security testing technique per AUTOSAR layer}
%	\label{fig:t_AUTOSAR}
%	\centering
%\end{figure*} 
%\begin{figure*}[h!]
%	\centering
%	\includegraphics[width=0.7\textwidth]{img/t_interface.png}
%	\caption{Security testing technique per AUTOSAR interface}
%	\label{fig:t_interface}
%	\centering
%\end{figure*}

\section{Discussion}
\label{sec:discussion}
% In this paper, we provide a systematic mapping study by analyzing the relation between security testing techniques, vehicle lifecycle phases, attacks and AUTOSAR layers and interfaces. 
This section presents the key findings of this study, security testing strategy for the automotive domain and potential threats to validity. 

\subsection{Key Findings \& Interpretation}
Our findings indicate that the most applied security testing techniques are penetration testing and dynamic analysis, and model-based testing. The model model-based testing is used to a high extent for testing access control policies as described in \cite{felderer2011classification}. With regards to AUTOSAR layers and functional interfaces, we identified that the services and application layer and multimedia/telematics interface are especially being taken into account. As for attacks and vehicle lifecycle phases, we discovered that attacks on privacy and availability, as well as design and engineering, production and operation by customer phases are considered for the most part. These are all related and demonstrate a strong co-occurrence between each other.

Likewise, our results show that the number of security testing approaches decreases starting from the initial design and engineering phase towards the final decommissioning phase. This relates to test efforts in the software development lifecycle (SDLC), where the focus is on the initial phases and especially on the design phase as presented in \cite{OWASP}. However, the key difference is that the last two phases of the vehicle lifecycle (maintenance and service, and decommissioning) are not addressed at all, which is not the case for the SDLC, where all the phases are addressed. According to \cite{bachmann}, it is crucial to consider security testing in each phase covering the whole development lifecycle, since various security issues may be identified during different stages.

Regarding the attacks, we observed that privacy and availability are addressed equally within the penetration testing and dynamic analysis, code-based and risk-based testing approaches. On the other hand, this is not the case with the model-based testing, where the privacy is addressed more compared to the availability. This is because many penetration testing techniques apply DDoS attacks to a large extent to conduct the testing, while model-based testing approaches do not. One may question the fact that we identified code-based techniques applying attacks on the privacy, availability and authentication. A reason might be that code-based techniques were paired together with penetration testing and dynamic analysis approaches as demonstrated in \citeR{bayer2015security} and \citeR{pike2017secure}.

In addition, we identified only a single security testing technique (penetration testing and dynamic analysis) addressing the runtime environment, microcontroller abstraction, complex drivers and microcontroller layers \citeR{groza2016evaluating}. We suspect that this may be the case because these layers are usually abstracted and are difficult to security test. Furthermore, researchers most likely prioritize testing other layers because they consider that the application and services layers provide more important functionalities, as well as a straight-forward access to the system.

As vehicle vendors are constantly improving their systems, they also tend to store more private data about their users. This can be dangerous since we discovered that no security testing approaches in maintenance and service and decommissioning phase of vehicle lifecycle are being used. Thus, it is not clear if something is being done with this data and if our data is really secure once the car goes into decommissioning process. Therefore, we identify this as one of the gaps that should be considered in the future.

Furthermore, we identified that there are no regression testing approaches, which is probably due to the issue that specific regression testing techniques for security are unfortunately still rare as indicated in \cite{Felderer2015}. This relates to no approaches addressing the maintenance and service phase of vehicle lifecycle. In other words, when a vehicle is brought for a regular service, the focus is on testing the performance and safety aspects. On the other hand, security is almost always neglected and it is usually an owner's task to make sure that his vehicle is still secure. However, we can assume that the security measures that were installed during the design phase may be disrupted throughout the operational phase. Apart from this, it is possible that security requirements changed over the time and require an update. Thus, it is necessary to consider security aspects during the maintenance phase and verify them using security regression testing techniques.

Finally, the safety interface \citeR{macher2014combined} of the AUTOSAR architecture is rarely tackled during security testing. There are various approaches that consider integration of security and safety in the automotive domain such as \cite{macher2017integrated}, \cite{DBLPBrunnerHSB17} and \cite{DBLPHuberBSCB18}. However, this is not the case with security testing and safety verification approaches. Therefore, new methods of combining security testing with approaches addressing safety aspects are needed. 

\subsection{Security testing strategy for the automotive domain}  
\begin{figure}[h!]
	\centering
	\includegraphics[width=0.8\columnwidth]{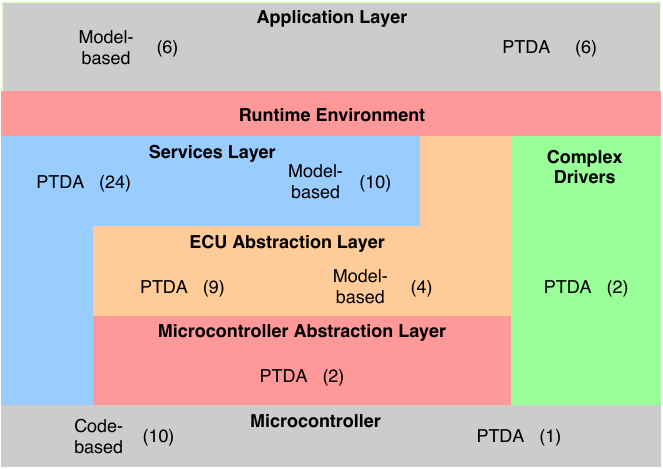}
	\caption{Security testing techniques per AUTOSAR Layer}
	\label{fig:autosar_sect}
	\centering
\end{figure}

Based on the results of our study, we are able to sketch a security testing strategy for the automotive domain. We do that by mapping the security testing techniques to the AUTOSAR layers. This is depicted in Figure \ref{fig:autosar_sect}, where each AUTOSAR layer contains security testing technique(s), excluding the runtime environment layer for which we did not identify a single testing technique because this layer is more abstracted and is usually out of testing scope compared to other layers. 

In doing so, we define a threshold value of 20\%, where each layer needs to be addressed using a security testing technique by at least 20\% of total number of selected publications. For example, penetration testing and dynamic analysis testing approach focusing on services layer of AUTOSAR architecture was identified in 24 out of 39 papers, which equals to 62\%. Therefore, we included this testing technique to Figure \ref{fig:autosar_sect}. As a result, we recommend testing the application, services and ECU abstraction layers using model-based and penetration testing and dynamic analysis as main techniques. In addition, the microcontroller layer should be primarily tested using code-based and, penetration testing and dynamic analysis techniques. Finally, the microcontroller abstraction layer and complex drivers layers should be tested using penetration testing and dynamic analysis security testing as main testing techniques. 

\subsection{Threats to validity}
During the research process, we considered possible threats to validity and tried to minimize them. Based on \cite{petersen2013worldviews}, we differentiate between descriptive validity, theoretical validity, generalizability, interpretive validity and repeatability.

\noindent
\textit{Descriptive validity}:
In order to minimize possible threats to descriptive validity, we created a classification table that was used to record the results. Hence, it is possible to apply this table at any time to record the same result set. In addition, each dimension within the table was taken from existing research publications. \textit{Theoretical validity}:
During the search process, it is possible that some papers were missed. For that purpose, we applied the keyword search together with the backward and forward snowballing. In addition, there is a probability that the selection and extraction process are biased. To counter that, we applied a cross-validation approach, where each publication was classified by at least two authors and in a case of any disagreements, a majority vote was taken and the papers were re-classified. Finally, it is possible that some activities within the analyzed approaches were misunderstood or overlooked. This relates in particular to AUTOSAR layers and functional interfaces, because the containing dimensions are very difficult to classify. The reason is that the information has to be taken out of the context from the selected publications. Thus, the researcher bias has been considered. \textit{Generalizability}: The research results in this study can only be generalized in the automotive domain. In addition, the results only apply to security testing techniques. Therefore, they are not applicable in any other domain. \textit{Interpretive validity}: The classification results were interpreted by all three authors. In doing so, we discussed any disagreement and applied statistical tools to analyze the results. However, there is a possibility that the final interpretation was impacted by a researcher bias. \textit{Repeatability}:
The process of this study was recorded in detail. This was described in the applied methodology (see Section \ref{sec:methodology}). In addition, we applied the existing guidelines. Therefore, it should be possible to repeat the process and conduct the equivalent study.

In addition, we should highlight the limitation of this study, which is that car vendors try to keep attack-related information to themselves to ensure that public knows little about security issues. However, innovative testing approaches are still published and most of the identified papers come from academia.

\section{Conclusion}
\label{sec:conclusion}
In this paper, we classified and analyzed security testing approaches applied in the automotive domain. In doing so, we conducted a systematic mapping study. In order to classify these approaches, we developed a classification table and investigated the following five dimensions: (1) security testing techniques, (2) AUTOSAR layers, (3) functional interfaces of AUTOSAR, (4) vehicle lifecycle phases and (5) attacks. In the next step, we classified the selected 39 publications based on the the aforementioned dimensions. The results showed the strong presence of penetration testing and dynamic analysis, and model-based testing approaches, which are used to address the application and services layer of the AUTOSAR architecture in design, production and operation phase of vehicle lifecycle. This is achieved by applying attacks focusing mostly to disrupt privacy and availability via the multimedia/telematics interface. Furthermore, we highlighted the need for regression testing approaches in the automotive domain, as well as the need for methods addressing security testing approaches together with the safety aspects. The future work consists of developing a security testing methodology, which will provide a detailed instructions on how the testing should be conducted in the automotive domain. 

%\end{document}  % This is where a 'short' article might terminate

\begin{acks}
  
    This work has been partially sponsored by the Austrian Ministry for Transport, Innovation and Technology (IKT der Zukunft, Project SALSA) 

\end{acks}

%\bibliographystyle{ACM-Reference-Format}
%\bibliography{sample-bibliography}
\bibliographystyle{ACM-Reference-Format}
\bibliography{automotive_security_testing}

\bibliographystyleR{ACM-Reference-Format}
\bibliographyR{R}

\end{document}